\documentclass[12pt]{article}
\usepackage{amsfonts,amssymb}
\usepackage{color}
\usepackage{epic}
\usepackage{graphics}

\def\lddots{\mathinner{\mkern1mu\raise1pt\hbox{.}\mkern2mu
\raise4pt\hbox{.}\mkern2mu\raise7pt\vbox{\kern7pt\hbox{.}}\mkern1mu}}
\makeatletter

\def\numberbysection{\@addtoreset{equation}{section}
\def\theequation{\thesection.\arabic{equation}}}
\makeatother
\numberbysection

\newcommand{\be}{\begin{eqnarray}}
\newcommand{\ee}{\end{eqnarray}}
\newcommand{\non}{\nonumber}

\begin{document}

\begin{titlepage}
\vskip 0.4cm \strut\hfill \vskip 0.8cm
\begin{center}


{\bf {\Large Boundary Lax pairs for the $A_{n}^{(1)}$ Toda
field theories}}

\vspace{10mm}

{\large {\bf Jean Avan\footnote{avan@u-cergy.fr}$^{a}$ and Anastasia
Doikou}\footnote{adoikou@upatras.gr}$^{b}$}

\vspace{10mm}

{\small $^a$ LPTM, Universite de Cergy-Pontoise (CNRS UMR 8089), Saint-Martin 2\\
2 avenue Adolphe Chauvin, F-95302 Cergy-Pontoise Cedex, France}

{\small $^b$ University of Patras, Department of Engineering
Sciences,\\ GR-26500 Patras, Greece}

\end{center}

\vspace{30mm}

\begin{abstract}
Based on the recent formulation of a general scheme to construct boundary Lax
pairs, we develop this systematic
construction  for the $A_n^{(1)}$ affine Toda field theories (ATFT). We work
out explicitly the first two models of the hierarchy, i.e. the
sine-Gordon ($A_1^{(1)}$) and the $A_2^{(1)}$ models. The
$A_2^{(1)}$ Toda theory is the first non-trivial example of the
hierarchy that exhibits two distinct types of boundary conditions.
We provide here novel expressions of boundary Lax pairs associated to both types of boundary conditions.

\end{abstract}

\vfill \baselineskip=16pt

\end{titlepage}

\section{Introduction}

The Lax representation of a classical dynamical system consists in the formulation of
two endomorphism-valued objects (in most cases matrix- or operator-valued) $L$ and
$M$, depending on the dynamical variables, such that the equations of motion are contained
in the iso-spectral evolution equation:
\be   {\partial}_t L = \Big[ L, M \Big] \label{dif0}
\ee
The Lax matrix $L$ therefore lives in some Lie algebra, finite-dimensional, loop
algebra, or differential algebra depending on the specific system. It is collectively called
``auxiliary algebra''.

The spectrum of $L$, or its associated monodromy matrix if $L$ is a differential operator,
provides therefore a generating set of natural time-invariant candidates to be identified as integrable Hamiltonians.
Liouville integrability of any dynamical system associated to a given function on this set
follows from the Poisson-commutation of the elements of the set, guaranteed by
the necessary and sufficient condition \cite{sts,bv} of the existence
of a generically linear Poisson structure
characterized by a so-called classical $r$-matrix, for the spectrum-generating
operator (Lax matrix or monodromy matrix).  A natural construction for $M$, given any specific
Hamiltonian built as a function $f$ on sp($L$), is then available in terms of $df$, $L$
and the $r$-matrix \cite{skl,sts}.

In our previous paper \cite{avandoikou} we have examined the situation when the Poisson structure
available for $L$ is expressed in terms of a non-dynamical classical $r$-matrix plus a set of non-dynamical parameters encapsulated in
a ``boundary'' matrix $K$ obeying some purely algebraic quadratic equation in terms of
this $r$-matrix. This more complicated equation (actually two distinct forms thereof)
represents a classical
version of the quantum Cherednik-Sklyanin reflection
algebra \cite{cherednik,sklyanin,kulishsklyanin} . The operator generating the Poisson
commuting Hamiltonians then combines $L$ and $K$. We have then defined a systematic construction
of the $M$ operator in terms of $r$, $L$ and $K$

Note here that the denomination of $K$ as ``boundary'' matrix reflects the fact that such parameters encode indeed boundary effects in the Hamiltonians when the matrix $L$ is a coproduct of local $l_i$ matrices on
a finite lattice $i= 1, \ldots, N$ or a monodromy of local differential operators on a finite line.
When by contrast $L$ is a purely local Lie-algebra valued matrix, such parameters may be better
characterized as coupling constants in a folding procedure (one is then considering a system on
a one-site space lattice for which the notion of a ``boundary'' has no sense).

Two fundamental situations were described in \cite{avandoikou}: linear Poisson structure and quadratic Poisson structure for the generating operator. The second one is relevant to describe systems
on a lattice or a continuous line. We shall here restrict ourselves to this
latter case, particularizing
it even more to degree-one differential operators $L = d/dx + l(x)$. The generating operator
is here the monodromy matrix of this differential operator computed between the two ends of the
finite line for $x$.

We consider here such operators for which the $r$-matrix is the one associated to the
generic affine $A_{n}^{(1)}$ Toda field theories. In this way we construct the associated Lax representation of ATFT with non-trivial integrable boundary conditions parametrized by the
$K$ matrix. Two ``boundary'' reflection equations must be considered, resp. characterized as ``soliton-preserving'' (SP) and ``soliton non-preserving'' (SNP) (see relevant studies at quantum and classical level \cite{menean}--\cite{doikouatft}).
More general reflection equation may be considered given a classical $r$-matrix, through the choice of some auxiliary-space anti-automorphism, but this extension shall be postponed for further studies.
It will be noted that this scheme
automatically yields boundary conditions compatible with
integrability. Careful evaluation of the contributions to the Hamiltonians
and the $M$ matrix at the edges of the $x$ line are required to get consistent
results.

Our essential purpose here is two-fold.
We shall first validate our general derivation by a comparison
of the $A_{1}^{(1)}$ and  Soliton-non-preserving $A_{2}^{(1)}$ cases with known results on low-dimension ATFT obtained by case-by-case analysis, basically for SNP boundary conditions \cite{durham}.
We shall indeed show that both derivations yield exactly identical formulations for the
boundary conditions for the dynamical fields and related Lax formulation
(with a proviso, related to analyticity
conditions, to be specified later). Similar consistency checks were also achieved in \cite{avandoikou} within the vector non-linear Schrodinger context.

We shall then provide novel expressions for boundary Hamiltonians and the associated Lax pairs in the yet untreated case of $A_{2}^{(1)}$ with SP boundary conditions, and we shall address some intriguing technical points that arise even in the simplest case, i.e. the sine-Gordon model.

\section{The general scheme}

Our analysis of classical integrable field theories with integrable boundary conditions relies
on the study of an associated auxiliary linear problem
(see \cite{ft} and references therein). Let us first recall the bulk case (no
extra ``boundary'' parameter) to fix the notations and recall the basic structures.

The Lax pair \cite{lax} formulation \cite{ZSh} of classical
integrable Hamiltonian systems consists in defining an auxiliary linear differential problem,
reading in the simplest case (order-1 differential operators):
\be && \Big ( {\partial}_x - {\mathbb U}(x,t, \lambda) \Big )  \Psi =0 \label{dif1}\\ &&
\Big ({\partial}_t - {\mathbb V}(x,t,\lambda)\Big ) \Psi =0
\label{dif2} \ee
${\mathbb U},\ {\mathbb V}$ are in general
$n \times n$ matrices with entries functions of the dynamical
fields, their space derivatives, and possibly the complex spectral parameter $\lambda$.
Compatibility conditions of the two differential equations
(\ref{dif1}), (\ref{dif2}) lead to the zero curvature condition
\cite{ ZSh, AKNS, abla} \be \partial_t{\mathbb U} -\partial_x {\mathbb V} + \Big
[{\mathbb U},\ {\mathbb V} \Big ]=0. \label{zecu} \ee The latter
equations give rise to the corresponding classical equations of
motion of the system under consideration. Natural conserved quantities are obtained from the monodromy matrix \be T(x,y, \lambda ) = {\cal P} \exp \{\int_x^y  {\mathbb U}(x',t, \lambda) dx' \}\ee once it is assumed that ${\mathbb U}$ obeys the classical linear Poisson algebraic relations:
\be \Big \{{\mathbb U}_a(x, \lambda),\ {\mathbb U}_b(y, \mu) \Big \} =\Big [r_{ab}(\lambda - \mu),\ {\mathbb U}_a(x, \lambda) +{\mathbb U}_b (y,\mu) \Big ]\
\delta(x-y), \label{ff0} \ee and consequently
$T(x,y,\lambda)$
satisfies (see \cite{ft}): \be \Big \{T_{a}(x,y,t,\lambda),\
T_{b}(x,y,t,\mu) \Big \}= \Big[r_{ab}(\lambda-\mu),~T_a(x,y,t,\lambda)T_b(x,y,t,\mu) \Big ]. \label{basic} \ee
One then immediately gets: \be \Big \{ tr T(\lambda),~ tr T(\mu) \Big \} =0 ~~~\mbox{where} ~~~~t(\lambda) = tr T(\lambda) \ee and thus $tr T(\lambda)$ yields the relevant conserved quantities.

We are now interested in implementing non-trivial integrable boundary conditions. We focus here
on two distinct types of integrable boundary conditions: the so-called soliton preserving (SP) and the soliton non-preserving (SNP). Formulation of the two distinct types of boundary conditions is
achieved by defining two types of Poisson structures for the modified monodromy matrices ${\cal T}$. These will in fact
represent the classical versions of the reflection
algebra ${\mathbb R}$, and the twisted Yangian ${\mathbb T}$ written in the
following forms (see e.g. \cite{sklyanin, maillet, mailfrei2, durham}):
\\
\\
{\bf (I) SP (reflection algebra)} \cite{cherednik, sklyanin} \be \Big \{{\cal T}_1(\lambda_1),\ {\cal
T}_2(\lambda_2) \Big \} &=& r_{12}(\lambda_1-\lambda_2){\cal
T}_{1}(\lambda_1){\cal T}_2(\lambda_2) -{\cal T}_1(\lambda_1)
{\cal T}_2(\lambda_2) r_{21}(\lambda_1 -\lambda_2) \non\\ & +&
{\cal T}_{1}(\lambda_1) r_{21}(\lambda_1+\lambda_2){\cal
T}_2(\lambda_2)- {\cal T}_{2}(\lambda_2)
r_{12}(\lambda_1+\lambda_2){\cal T}_1(\lambda_1) \label{refc1} \ee
\\
{\bf (II) SNP (($q$)-twisted Yangian)} \cite{molev, moras} \be \Big \{{\cal T}_1(\lambda_1),\ {\cal
T}_2(\lambda_2) \Big \} &=& r_{12}(\lambda_1-\lambda_2){\cal
T}_{1}(\lambda_1){\cal T}_2(\lambda_2) -{\cal T}_1(\lambda_1)
{\cal T}_2(\lambda_2) r^{t_1 t_2}_{21}(\lambda_1 -\lambda_2) \non\\
& +& {\cal T}_{1}(\lambda_1)
r_{12}^{t_1}(\lambda_1+\lambda_2){\cal T}_2(\lambda_2)- {\cal
T}_{2}(\lambda_2) r_{21}^{t_2}(\lambda_1+\lambda_2){\cal
T}_1(\lambda_1) \label{refc2} \ee
In most well known physical cases, such as the $A^{(1)}_{{\cal N}
-1}$ $r$-matrices $r_{12}^{t_1 t_2} = r_{21}$, hence all the expressions above may
be written in a simpler form.

In order to construct representations of (\ref{refc1}), (\ref{refc2}) yielding the
generating function of Poisson-commuting Hamiltonians realizing
the integrals of motion for a new classical integrable system, one now introduces non-dynamical
representations ($K^{\pm}$) of the algebra ${\mathbb
R}$ (${\mathbb T}$). The non-dynamical condition: \be \Big \{
K_1^{\pm}(\lambda_1),\ K_2^{\pm}(\lambda_2) \Big \}=0 \label{kso}
\ee transforms (\ref{refc1}), (\ref{refc2}) into algebraic equations for
$K^{\pm}$. We consider then any bulk monodromy matrix $T$ with Poisson
structure (\ref{basic})
and we define in addition: \be \hat T(\lambda)= T^{-1}(-\lambda) ~~~\mbox{for SP},
~~~~~\hat T(\lambda) =T^{t}(-\lambda) ~~~\mbox{for SNP}. \label{notation} \ee
One expects that a more self-contained formulation may involve a more
general anti-automorphism of the auxiliary algebra where $T$ lives. A
corresponding reformulation of the relevant Poisson structures should
also be proposed in that framework. We shall leave this generalization
for later studies.

Generalized ``monodromy'' matrices,
realizing the corresponding
algebras ${\mathbb R},\ {\mathbb T}$, are finally given by the following
expressions \cite{sklyanin, durham}: \be && {\cal T}(x,y,t,\lambda) = T(x,y,t,\lambda)\ K^{-}(\lambda)\ \hat T(x,y,t,\lambda).
\label{reps0} \ee The generating function of the involutive
quantities is defined as \be t(x,y,t,\lambda)= tr\{
K^{+}(\lambda)\ {\cal T}(x,y,t,\lambda)\}. \label{tr} \ee Indeed
one shows: \be  \Big \{t(x,y,t,\lambda_1),\ t(x,y,t,\lambda_2)
\Big \} =0, ~~~ \lambda_1,\ \lambda_2 \in {\mathbb C}.
\label{bint0} \ee

The systematic Lax formulation in the case of open boundary conditions
is described in \cite{avandoikou}. More precisely it was shown in \cite{avandoikou} that
\be \Big \{ \ln\ t(\lambda),\ {\mathbb U}(x, \mu) \Big \} = {\partial {\mathbb V}(x,\lambda, \mu) \over
\partial x} + \Big [ {\mathbb V}(x,\lambda, \mu),\ {\mathbb U}(x, \mu) \Big ] \label{defin} \ee where we define:
\\
\\ {\bf (I) SP}
\be {\mathbb V}(x, \lambda, \mu) &=& t^{-1}(\lambda) \ tr_a \Big (K_a^+(\lambda) T_a(0, x, \lambda) r_{ab}(\lambda -\mu) T_a(x,
-L, \lambda) K_a^-(\lambda) T_{a}^{-1}(0, -L, -\lambda) \non\\ &+& K_a^+(\lambda)T_a(0,
-L, \lambda) K_a^-(\lambda)  T_a^{-1}(x, -L, -\lambda)
r_{ba}(\lambda +\mu) T_a^{-1}(0, x, \lambda) \Big ),
 \label{final1} \ee  \\
{\bf (II) SNP}
 \be {\mathbb V}(x,\lambda,
\mu) &=& t^{-1}(\lambda) \ tr_a \Big ( K_a^+(\lambda)  T_a(0, x, \lambda) r_{ab}(\lambda -\mu) T_a(x,
-L, \lambda) K_a^-(\lambda) T_{a}^{t_a}(0, -L, -\lambda) \non\\ &+& K_a^+(\lambda)T_a(0,
-L, \lambda) K_a^-(\lambda)  T_a^{t_a}(x, -L, -\lambda)
r_{ab}^{t_a}(-\lambda -\mu) T_a^{t_a}(0, x, -\lambda)\Big ),
\label{final2} \ee  \\
Particular attention should be paid to the boundary points $x =0,\
-L$. Indeed, for these two points one has to take into
account that $T(x,x ,\lambda) = \hat T(x,x, \lambda) ={\mathbb I}$.
\\
\\ {\bf (I) SP}
\be
{\mathbb V}(0,t, \lambda, \mu) &=& t^{-1}(\lambda)\  tr_a \Big (K_a^+(\lambda) r_{ab}(\lambda-\mu) T_a(0, -L, \lambda)
K_a^-(\lambda) T_a^{-1} (0, -L, -\lambda) \non\\ &+& K_a^{+}(\lambda)T_a(0, -L, \lambda) K_a^-(\lambda) T_a^{-1}(0, -L, -\lambda)
r_{ba}(\lambda + \mu) \Big ) \non\\
{\mathbb V}(-L,t, \lambda,\mu) &=& t^{-1}(\lambda)\  tr_a \Big (K_a^+(\lambda)T_a(0, -L, \lambda) r_{ab}(\lambda-\mu)K^-(\lambda)
T_a^{-1} (0, -L, -\lambda) \non\\ &+&  K_a^{+}(\lambda)T_a(0, -L, \lambda) K_a^-(\lambda) r_{ba}(\lambda+\mu)T_a^{-1}(0, -L, -\lambda)
 \Big )
\label{boundary} \non\\ \ee
\\ {\bf (II) SNP}
\be
{\mathbb V}(0,t, \lambda, \mu) &=& t^{-1}(\lambda)\  tr_a \Big (K_a^+(\lambda) r_{ab}(\lambda-\mu) T_a(0, -L, \lambda)K_a^-(\lambda)
T_a^{t_a} (0, -L,-\lambda) \non\\ &+&  K_a^{+}(\lambda) T_a(0, -L, \lambda) K_a^-(\lambda) T_a^{t_a}(0, -L, -\lambda)
r_{ab}^{t_a}(-\lambda-\mu) \Big ) \non\\
{\mathbb V}(-L,t, \lambda,\mu) &=& t^{-1}(\lambda)\  tr_a \Big (K_a^+(\lambda)T_a(0, -L, \lambda) r_{ab}(\lambda-\mu)K^-(\lambda)
T_a^{t_a} (0, -L, -\lambda) \non\\ &+&  K_a^{+}(\lambda)T_a(0, -L,\lambda) K_a^-(\lambda) r_{ab}^{t_a}(-\lambda-\mu)T_a^{t_a}(0, -L, -\lambda)
 \Big )
\label{boundary2} \non\\ \ee
Notice that all the boundary information, incorporated in $K^{\pm}$, appears only at the
boundary points $x=0, ~-L$. We shall presently see that the bulk expression has in fact no
dependence whatsoever on the reflection $K$ matrix, since it is canceled by the $t^{-1}(\lambda)$ factor.
Note finally that the expressions derived in (\ref{final1})--(\ref{boundary}) are
universal, that is independent of the choice of model.

\section{The $A_{n}^{(1)}$ ATFT: brief review}

We are now in a position to systematically construct the Lax
representation for any extension of the $A_{n}^{(1)}$ ATFT following the
scheme defined above, given any solution $K$ of the algebraic boundary
equations defined by the $A_{n}^{(1)}$ Toda classical $r$-matrix. Note that the associated  boundary Hamiltonians have been extracted through the asymptotic expansion of the open transfer matrix in \cite{macnt} for sine-Gordon and in \cite{doikouatft} for the $A_{2}^{(1)}$ ATFT.
We shall consider in the following sections two particular
examples, that is the prototype model of the hierarchy, i.e. the
sine-Gordon model, as well as the $A_2^{(1)}$ case. The $A_2^{(1)}$
model is indeed the first non-trivial example of this set that may exhibit
both types of boundary conditions. It is worth noting that in sine-Gordon
the two boundary conditions coincide due to the fact that the model
is self dual.

Recall first the classical $r$-matrix associated to the generic $A_{n}^{(1)}$ affine
Toda field theory in particular is given by\footnote{Notice that
the $r$-matrix employed here is in fact $r_{12}^{t_1 t_2}$ with
$r_{12}$ being the matrix used e.g. in \cite{done, doikounp}}
\cite{jimbo} \be r(\lambda) = {\cosh(\lambda) \over \sinh
(\lambda)} \sum_{i=1}^{n+1} e_{ii}\otimes e_{ii} + {1\over \sinh
(\lambda)} \sum_{i \neq j =1}^{n+1} e^{[sgn(i-j) -(i-j) {2 \over
n+1}  ] \lambda } e_{ij} \otimes e_{ji}. \label{rc} \ee with
$(e_{ij})_{kl} \equiv \delta_{ik}\ \delta_{jl}$. Note that
the classical $r$-matrix (\ref{rc}) is written in the so-called
principal gradation as in \cite{durham, gand} (see details on the gauge transformation changing the principal to the homogeneous gradation in
\cite{doikouatft}).
We recall the Lax pair for a generic $A_{n}^{(1)}$
theory \cite{olive}: \be && {\mathbb V}(x,t,u) = - {\beta \over 2}\
\partial_x \Phi \cdot H + {m \over 4}\ \Big (u\ e^{{\beta \over 2}
\Phi \cdot H}\ E_{+}\ e^{- {\beta \over 2} \Phi \cdot H} -{1\over
u}\ e^{-{\beta \over 2}
\Phi \cdot H}\ E_{-}\ e^{{\beta \over 2} \Phi \cdot H}\Big) \non\\
&& {\mathbb U}(x,t,u) = {\beta \over 2}\ \Pi \cdot H + {m \over
4}\ \Big (u\ e^{{\beta \over 2} \Phi \cdot H}\ E_{+}\ e^{- {\beta
\over 2} \Phi \cdot H} +{1\over u}\ e^{-{\beta \over 2} \Phi \cdot
H}\ E_{-}\ e^{{\beta \over 2} \Phi \cdot H}\Big)  \label{lpair}
\ee $\Phi,\ \Pi$ are conjugated $n$-vector fields, with components $\phi_i,\
\pi_i,\ i \in \{1, \ldots , n\}$, $~u=e^{{2\lambda \over n+1}}$ is
the multiplicative spectral parameter. To compare with the
notation used in \cite{durham} we set ${m^2 \over 16}= {\tilde m^2
\over 8}$ ($\tilde m$ denotes the mass in \cite{durham}). Note
that eventually in \cite{durham} both $\beta,\ \tilde m$ are set
equal to unit.

We also define: \be E_+ = \sum_{i=1}^{n+1} E_{\alpha_i},
~~~~~E_{-}=\sum_{i=1}^{n+1} E_{-\alpha_i} \ee $\alpha_i$ are the
simple roots plus the extended (affine) root, $H$ ($n$-vector) and $E_{\pm \alpha_i}$ are the
algebra generators in the Cartan-Weyl basis, and they satisfy the Lie algebra relations: \be &&
\Big [H,\ E_{\pm \alpha_i} \Big]= \pm \alpha_i E_{\pm \alpha_i},
\non\\ && \Big [E_{\alpha_i},\ E_{-\alpha_i} \Big ] = {2\over
\alpha_i^2}\ \alpha_i \cdot H \ee Explicit expressions on the
simple roots and the Cartan generators are presented below. Notice
that the Lax pair has the following behavior: \be
{\mathbb V}^t(x, t, -u^{-1}) = {\mathbb V}(x,t, u), ~~~~~{\mathbb
U}^t(x, t, u^{-1}) = {\mathbb U}(x,t,u) \ee where $^t$ denotes
usual transposition.

We provide below explicit
expressions of the simple roots and the Cartan generators for
$A_{n}^{(1)}$ \cite{georgi}. The vectors $\alpha_{i} =(\alpha_{i}^{1} \,, \ldots \,, \alpha_{i}^{n})$ are the simple
roots of the Lie algebra of rank $n$ normalized to unity
$\alpha_{i} \cdot \alpha_{i} = 1$, i.e. \be \alpha_{i} = \Bigl(0
\,,  \ldots \,, 0 \,, -\sqrt{i-1\over 2i} \,,
\stackrel{\stackrel{i^{th}}{\downarrow}} {\sqrt{i+1\over 2i}} \,,
0 \,,  \ldots \,, 0 \Bigr), ~~~~i \in \{1, \ldots n \} \ee
The fundamental weights $\mu_{k} = (\mu_{k}^{1} \,, \ldots
\,, \mu_{k}^{n}) \,, \quad  k = 1 \,, \ldots \,, n $ are defined as (see,
e.g., \cite{georgi}). \be \alpha_{j} \cdot \mu_{k} = {1\over 2}
\delta_{j,k} \,. \label{important} \ee The extended (affine) root
$a_{n+1}$  is provided by the relation \be \sum_{i=1}^{n+1} a_i
=0. \ee The Cartan-Weyl generators in the defining
representation are: \be E_{\alpha_{i}} &=& e_{i\ i+1} \,, \qquad
E_{-\alpha_{i}} = e_{i+1\ i} \,, \qquad E_{\alpha_n} = - e_{n+1\
1} \,, \qquad E_{-\alpha_n} = - e_{1\ n+1}\non \\
H_{i} &=& \sum_{j=1}^{n} \mu_{j}^{i} (e_{j j} -e_{j+1\ j+1}) \,,
\qquad i = 1 \,, \ldots \,, n. \label{cartan/weyl/basis} \ee

\section{The boundary $A_{1}^{(1)}$ case: sine-Gordon model}

Let us rewrite the Lax operator for the bulk sine Gordon
model\footnote{To recover the generic form (\ref{lpair}) from  (\ref{laxsg}) we consider the
following identifications \be {\beta \over i} \to  \beta,
~~~~\phi \to - \phi, ~~~~u\to -u \label{id0}\ee Here (\ref{laxsg}) we clearly
consider the sine-Gordon, however after
implementing identifications (\ref{id0}) we obtain the sinh-Gordon model.}, \be
{\mathbb U}(x,t,u) = {\beta \over 4i} \pi(x) \sigma_3 + {mu \over
4i}e^{{i \beta \over 4}\phi \sigma_3} \sigma_2 e^{-{i \beta \over
4}\phi \sigma_3}-{mu^{-1}\over 4i}e^{-{i \beta \over 4}\phi
\sigma_3} \sigma_2 e^{{i \beta \over 4 }\phi \sigma_3} \label{laxsg} \ee
$\sigma_{i}$ are the 2-dimensional Pauli matrices.

Bearing in mind the expression for ${\cal T}$ (\ref{reps0}) it is clear that
we need to consider the formal series expansion of $T$ and $T^{-1}(u^{-1})$. But from the
following symmetry of the Lax operator: \be {\mathbb U}(u^{-1},
\phi, \pi) = {\mathbb U} (-u. -\phi, \pi)\ee we see that \be
T(u^{-1}, \phi, \pi) = T(-u, -\phi, \pi). \ee We aim at expressing the term of order $u$ in ${\mathbb U}$
independently of the fields, after applying a suitable gauge
transformation \cite{ft}. More precisely,
consider the following gauge transformation such that \be T(x,y,u)
= \Omega(x)\ \tilde T(x,y, u)\ \Omega^{-1}(y), \non\\ \Omega(x) =\mbox{diag} \Big (\Omega_1(x),\ \Omega_2(x) \Big )= e^{{i\over 4}\beta \phi(x) \sigma_3 } \ee then the gauge
transformed operator $\tilde {\mathbb U}$ is expressed as: \be
\tilde {\mathbb U}(x,t,u)=  {\beta \over 4i} \mathfrak{f}(x) \sigma_3 +
{mu \over 4i} \sigma_2 -{mu^{-1}\over 4i}e^{-{i \beta \over 2}\phi
\sigma_3} \sigma_2 e^{{i \beta \over 2 }\phi \sigma_3}\ee where we define
\be  \mathfrak{f}(x,t) = \pi(x,t) + \phi'(x,t). \ee

Let $T'(u) = T(u^{-1})$ then we introduce the following decomposition
for $\tilde T$, $~\tilde T'$ as $|u| \to \infty$ \cite{ft} \be &&
\tilde T(x,y,u) = ({\mathbb I}
+W(x, u))\ \exp[Z(x,y,u)]\ ({\mathbb I} +W(y,u))^{-1}, \non\\
&&\tilde T'(x,y, u) = ({\mathbb I}+ \hat W(x, u))\ \exp[\hat
Z(x,y,u)]\ ({\mathbb I} + \hat W(y,u))^{-1}, \label{exp0} \ee where
the hat simply denotes that $u\to -u,\ \phi \to -\phi$. $W,\ \hat W$
are off diagonal matrices and $Z,\ \hat Z$ are purely diagonal. Also
\be Z(u) = \sum_{k=-1}^{\infty} {Z^{(k)} \over u^{k}}, ~~~~W(u)= \sum_{k=0}^{\infty}{W^{(k)} \over u^k}. \label{expa} \ee Inserting
the latter expressions (\ref{expa}) in (\ref{dif1}) one may identify
the matrices $W^{(k)}$ and $Z^{(k)}$. Indeed, from equation (\ref{dif1}) we conclude that the gauge transformed
operators satisfy: \be && {d Z \over dx} = \tilde {\mathbb U}^{(D)}
+ \tilde {\mathbb U}^{(O)} W \non\\ && {d W \over d x} + W \tilde
{\mathbb U}^{(D)} - \tilde {\mathbb U}^{(D)}W + W \tilde {\mathbb
U}^{(O)} W -\tilde {\mathbb U}^{(O)}=0 \label{basic0}\ee where
$\tilde {\mathbb U}^{(D)},\ \tilde {\mathbb U}^{(O)}$ are the
diagonal and off diagonal parts of $\tilde {\mathbb U}$
respectively. By solving the latter equations we may identify the
matrices $Z,\ W$.  It is sufficient for our purposes here to
identify only the first couple of terms of the expansions. Indeed
based on equation (\ref{basic0}) we conclude (see also \cite{ft}):
\be W^{(0)} = i \sigma_1, ~~~~W^{(1)} = -{i \beta \over  m}
\mathfrak{f}(x) \sigma_1 \ee
Note that the leading contribution as $i u \to \infty$ comes
from the $e^{Z_{22}}$ term. We assumed here for simplicity, but
without losing generality, Schwartz boundary conditions at the end
point $x =-L$, that is $\pi(-L)=\phi(-L) =0$ and $K^- \propto
{\mathbb I}$. We may rewrite the expression for the boundary
operator ${\mathbb V}$ (let $\hat r_{ab} = r_{ba}$)\footnote{Note that in this particular case
the SP and SNP boundary conditions coincide because: \be
r_{12}(\lambda) = V_1\ r^{t_1}_{12}(-\lambda)\ V_1, ~~~~~V= antidiag(1,\ 1)\ee} as: \be && {\mathbb V}(x,t, u, v)= t^{-1}(u)e^{Z_{22} -\hat Z_{22}} \Big ((1+\hat W(0))^{-1}
\Omega(0) K^+(u) \Omega(0) (1+W(0)) \Big )_{22} \non\\ && \Big \{
\Big [ (1+W(x))^{-1}\Omega^{-1}(x) r(uv^{-1}) \Omega(x) (1
+W(x))\Big ]_{22} \non\\&& +\Big [ (1+\hat W(x))^{-1}\Omega(x)
\hat r(u v) \Omega^{-1}(x) (1 +\hat W(x))\Big ]_{22} \Big
\} \non\\ \ee but it is easy to show for the transfer matrix
(\ref{tr}) as $|u| \to \infty$: \be t(u) = e^{Z_{22} -\hat Z_{22}} \Big
((1+\hat W(0))^{-1} \Omega(0) K^+(u) \Omega(0) (1+W(0)) \Big
)_{22} \ee and finally \be {\mathbb V}(x,t,u,v) &=& \Big [
(1+W(x))^{-1}\Omega^{-1}(x) r(uv^{-1}) \Omega(x) (1 +W(x))\Big
]_{22} \non\\ &+& \Big [ (1+\hat W(x))^{-1}\Omega(x)
\hat r(u v) \Omega^{-1}(x) (1 +\hat W(x))\Big ]_{22}.
\label{bulk}\ee

Again using the ansatz for the monodromy matrix we obtain from
(\ref{boundary}) for the end point $x=0$: \be && {\mathbb V}(0,t,u,v)= \Big [ (1+\hat
W(0))^{-1}\Omega(0) K^+(u) \Omega(0) (1+W(0)) \Big ]^{-1}
\non\\ && \Big \{\Big [(1+\hat W(0))^{-1} \Omega(0) K^+(u) r(uv^{-1})
\Omega(0) (1+W(0))\Big ]_{22} \non\\
&& +  \Big [(1+\hat W(0))^{-1} \Omega(0) \hat r(u v) K^+(u)
\Omega(0) (1+W(0))\Big ]_{22}\Big \}. \label{boundary22}\ee
The $r$-matrix is given in (\ref{rc}) and we consider below two cases with non-diagonal and diagonal $K$-matrix respectively.

\subsection{Non-diagonal $K$-matrix}

We shall first examine the case with the generic non-diagonal $K$-matrix \cite{GZ, dvg}
\be K^+(\lambda) ={1 \over \kappa} \sinh (\lambda + i \xi )e_{11} + {1 \over \kappa}\sinh(-\lambda + i \xi ) e_{22} + x^+ \sinh (2 \lambda)e_{12}  + x^-\sinh (2 \lambda)e_{21} \label{kmatrix0} \ee $\xi,\ x^{\pm}$ are a priori free independent
 boundary parameters.
The next step is to expand expressions (\ref{bulk}),
(\ref{boundary22}) in powers of $u^{-1}$, and identify the first
order term of the expansion. Taking into account the expansion of
$W$ as well as ($|u| \to \infty $) \be && K^+(u) \sim K^{+(0)} + u^{-1}K^{+(1)} + {\cal
O}(u^{-2}), ~~~~r(uv^{-1}) \sim r^{(0)} + u^{-1} r^{(1)} + {\cal
O}(u^{-2}), \non\\ && \hat r(uv) \sim r^{(0)} + u^{-1}\hat
r^{(1)} + {\cal O}(u^{-2}) \label{kmatrix}\ee where we define: \be && K^{+(0)} = x^+ e _{12} + x^-e_{21},
~~~~K^{+(1)} = {e^{i\xi} \over \kappa }e_{11}  -{e^{-i\xi} \over \kappa }e_{22}, \non\\
&& r^{(0)} = \sum_{i=1}^{2}e_{ii}\otimes e_{ii}, ~~~~~r^{(1)} = 2v
(e_{12}\otimes e_{21}+e_{21}\otimes e_{12}), ~~~~\hat r^{(1)} = 2
v^{-1} (e_{12} \otimes e_{21} + e_{21} \otimes e_{12}) \non\\ \label{kmatrix1} \ee
we may expand ${\mathbb V}(u, v)$ in powers of
$u^{-1}$. Multiplying the resulting expression by a factor ${m \over 4 i }$ we obtain at first order: \be {\mathbb V}(x,t,v) = {\beta \over 4i}
\phi'(x,t)\sigma_3 + {v m \over 4i}\ \Omega(x,t)\ \sigma_2\
\Omega^{-1}(x,t) + {v^{-1}m \over 4i} \Omega^{-1}(x,t)\ \sigma_2\
\Omega(x,t) \non\\ \label{bulk1} \ee We see that the operator ${\mathbb
V}(x,t,v)$ at any point $x \neq 0$ coincides with the bulk operator
(\ref{lpair}), consistently with the fact that $\cal H$ coincides with the bulk sine-Gordon  boundary Hamiltonian \cite{GZ, macnt}  except for $x=0$:
\be {\cal H} = \int_{-L}^0 dx \Big [{1\over 2} (\pi^2+ \phi^{'2}) + {m^2 \over \beta^2}
(1-\cos\beta \phi) \Big ] + {4 P m \over \beta^2} \cos{\beta
\phi(0) \over 2}  -{4 Q m\over \beta^2} \sin {\beta \phi(0) \over
2} \label{hsg}\ee Recall that in \cite{macnt} the latter Hamiltonian was obtained as the first order term from the expansion of the generating function $t(u)$ as $|u| \to \infty$, assuming Schwartz boundary conditions at $x=-L$. The boundary parameters $P,\ Q$ are
related to the parameters $\xi,~ \kappa$ of the $K$ matrix as: \be P = {
e^{i\xi} - e^{-i\xi} \over 4 \kappa}, ~~~~~~~Q = {e^{i \xi} +
e^{-i\xi} \over 4 \kappa i}. \label{ident} \ee We should stress that the
constraint $x^+ = -x^-$ here arises by requiring that the expansions of the transfer matrix as $iu \to \infty$ and $iu \to -\infty$ provide the same Hamiltonians (again, an analyticity condition at infinity). Such a requirement leads also to the cancelation of boundary terms proportional to $\phi'(0)$.

It is clear that the bulk ${\mathbb V}$-operator is independent of the choice of $K$-matrix. Expanding carefully the boundary expression (\ref{boundary2}) and
multiplying the result with a factor ${m \over 4 i}$ we obtain at the boundary point:
\be && {\mathbb V}^{(b)}(0,t,v) = {\mathbb V}(0,t,v) + \Delta {\mathbb V}(0,t,v), ~~~~~\mbox{where} \non\\
&& \Delta {\mathbb V}(0,t,v) =  - {\beta \over
4i}\phi'(0)\sigma_3 - {m \over 8} \Big ({e^{i \xi} \over 2
\kappa } e^{{i\beta \over 2}\phi(0)} + {e^{-i\xi}\over
2\kappa}e^{-{i\beta \over 2}\phi(0)} \Big )\sigma_3 \ee ${\mathbb V}(0,t,v)$ is provided by the bulk expression (\ref{bulk1}). Note that
all the boundary information is incorporated at the boundary point
$x=0$. The equations of motion and the corresponding boundary
conditions emerge in this Lax formulation from the zero curvature condition. The zero curvature condition for the `bulk' Lax pair yields the familiar equations of motion for the sine-Gordon model. We should note that analyticity requirements on the boundary Lax pair leads to extra constraints among the boundary parameters, i.e. $x^{+} = - x^{-}$ (see also \cite{macnt}). The boundary operator found here is associated to the boundary Hamiltonian (\ref{hsg}).
Note that a particular choice of diagonal $K^+$ matrix leads to discrepancies between the two descriptions (Hamiltonian vs Lax pair). This suggests that one has to consider as a starting point a generic solution of the reflection equation with several boundary parameters, which may then satisfy further constraints dictated by certain consistency requirements.

The relevant boundary conditions are obtained by considering
the zero curvature condition (\ref{zecu}) at the point $x=0$:
\be && \dot{{\mathbb U}}(0,t,v) - {d \over
d x }{\mathbb V}^{(b)}(x,t,v)\vert_{x=0} + \Big [{\mathbb U}(0,t,v),\
{\mathbb V}^{(b)}(0,t,v) \Big ]=0 \Rightarrow \\
&& \dot{{\mathbb U}}(0,t,v) - \lim_{\delta \to 0} {{\mathbb
V}(\delta,t,v) -{\mathbb V}(0,t,v)-\Delta {\mathbb V}(0,t,v) \over
\delta} + \Big [{\mathbb U}(0,t,v),\ {\mathbb V}(0,t,v) + \Delta
{\mathbb V}(0,t,v) \Big ]=0. \non\\ \label{vx}\ee Explicit expression of the derivative of
${\mathbb V}$  at $x=0$ in (\ref{vx}) indicates \be \Delta {\mathbb V}(0,t,v) =0, \ee
in order to eliminate a potential uncompensated divergence due to $\Delta{\mathbb V}$.

Finally from the `bulk' zero curvature condition and from
the later expression the following equations of motion and mixed boundary conditions are
entailed: \be && \ddot{\phi}(x, t) - \phi''(x, t) = -{m^2 \over \beta} \sin(\beta \phi(x, t))
\non\\ && \beta \phi'(0) = {m \over 2 i \kappa} \cos(\xi + {\beta
\over 2} \phi(0)), \label{bc1} \ee which of course coincide with the equations of
motion found in \cite{GZ, macnt} (recall also the identification of
boundary parameters (\ref{ident})).
In a consistent way the boundary conditions are obtained exactly from the Hamiltonian through \be  {\partial \phi(x, t) \over \partial t} = \{{\cal H}, ~\phi(x,t) \}, ~~~~~~~{\partial \pi(x, t) \over \partial t} = \{{\cal H}, ~\pi(x,t) \}, ~~~~x\in [-L,~ 0]
\ee by noticing that the contribution containing the term $\phi(0)$ in the Hamiltonian yields a $\delta(0)$ term in the equations of motion for $\pi(x)$ since $\{\phi(x), ~ \pi(y) \}= \delta(x-y)$. Elimination of this term yields exactly the boundary conditions (\ref{bc1}).

Requiring cancelation of the $\Delta {\mathbb V}(0)$ term is equivalent to requiring that the formal series expansion in $u^{-1}$ coincides for ${\mathbb V}(0,t, u, v)$ and ${\mathbb V}(x,t, u,v)$ at $x \to 0$. Indeed the technical origin of $\Delta {\mathbb V}(x=0)$ is the non-commutation of limits $x \to 0$ and $u \to \infty$, in particular in $e^Z$. If these limits are required to commute then ${\mathbb V}(x,t, u)$ has its analytic behavior in $x, ~u$ continued to the limit $x=0$, which may suggest that it can be analytically continued ``beyond'' the boundary. This may in turn be a relevant consistency condition in implementing the notion of ``gluing'' different boundary systems. We have thus established a straightforward and elegant way to extract the associated boundary conditions from the zero curvature condition.

\subsection{Diagonal $K$-matrix}

We shall now consider the diagonal $K$-matrix \cite{GZ, dvg}
\be K^+(\lambda) =\sinh (\lambda + i \xi )e_{11} + \sinh (-\lambda + i \xi ) e_{22}.  \label{kmatrix01}\ee In particular we shall be mostly interested in the degenerate limit where $i \xi \to \infty$.

The relevant boundary Hamiltonian, obtained from the first order term of the expansion of $\ln t(\lambda)$, is given by:
\be {\cal H} = \int_{-L}^0 dx \Big [{1\over 2} (\pi^2+ \phi^{'2}) + {m^2 \over \beta^2} (1-\cos\beta \phi) \Big ] + {2 \over \beta } \phi'(0) {\cos(\xi+{\beta \over 2} \phi(0)) \over \sin(\xi + {\beta \over 2}\phi(0)) }. \label{hsg1}\ee The boundary contribution in (\ref{hsg1}) is not identical with the $\kappa =0$ limit of the boundary conditions in (\ref{hsg}). One needs to normalize the $K$-matrix as $u K(\kappa \to 0)$ to get a consistent $u$-expansion, hence ${\cal H}$ in (\ref{hsg1}) picks boundary contribution from higher orders in (\ref{kmatrix}).
By requiring the boundary term, proportional to $\phi'(0)$ to disappear we obtain the following constraint \be \cos(\xi+{\beta \over 2} \phi(0))=0, \ee which of course may be seen as the boundary condition to the associated equations of motion, as we shall see below.

The next step as in the previous case is to expand expressions (\ref{bulk}), (\ref{boundary22}) in powers of $u^{-1}$, and identify the associated ${\mathbb V}$-operator from the first
order term of the expansion. As we have seen the bulk ${\mathbb V}(u, v)$-operator is independent of the choice of boundary conditions, i.e. the $K$-matrix and is given at any point $x \neq 0$ by
(\ref{lpair}). Expanding carefully the boundary expression (\ref{boundary2}) and
multiplying the result with a factor ${-m \over 2 i}$ we obtain at the boundary point:
\be && {\mathbb V}^{(b)}(0,t,v) = \non\\ && {\beta \over
i\Delta^2} y^+ y^- \Omega_1^2(0) \Omega_2^2(0) \phi'(0)\sigma_3 + {m  \over 2 \Delta}\Omega_1(0) \Omega_2(0)  \Big ( v(y^+ e_{21} - y^- e_{12}) + v^{-1}  ( y^- e_{21} - y^+ e_{12}) \Big ) \non\\ \ee where $\Delta = y^+ \Omega_1^2(0) + y^- \Omega_2^2(0)$, and $y^{\pm} = \pm e^{\pm i\xi}$.

By requiring ${\mathbb V}(0) = {\mathbb V}^{(b)}(0)$ we obtain the corresponding boundary conditions.
Finally from the `bulk' zero curvature condition and from
the later expression the following equations of motion and mixed boundary conditions are
entailed: \be && \ddot{\phi}(x, t) - \phi''(x, t) = -{m^2 \over \beta} \sin(\beta \phi(x, t))
\non\\ && \cos(\xi + {\beta \over 2}\phi(0))=0, \label{bc2}\ee which of course coincide with the equations of motion found in \cite{GZ, macnt} (recall also the identification of
boundary parameters (\ref{ident}) and the boundary conditions found earlier. Notice that the obtained boundary conditions are easily obtained from the generic situation described in (\ref{bc1}) by simply setting the non diagonal contributions to zero.

We are however mostly interested in the case where the $K$-matrix is degenerate (in the homogeneous gradation). Consider for instance the situation where $K(\lambda) = \mbox{diag}(e^{\lambda},\ e^{-\lambda})$. The Hamiltonian and boundary ${\mathbb V}$-operator in this case are given respectively by:
\be {\cal H} = \int_{-L}^0 dx \Big [{1\over 2} (\pi^2+ \phi^{'2}) + {m^2 \over \beta^2} (1-\cos\beta \phi) \Big ] + {2 \over \beta } \phi'(0) \ee and \be {\mathbb V}^{(b)}(0,v) =  {m\over 4} \Omega_1(0) \Omega_2^{-1}(0) \Big ( ve_{21} - v^{-1}e_{12} \Big ). \ee
The boundary conditions emerging from the Hamiltonian are: $\phi'(0) =0$ whereas requiring ${\mathbb V}^{(b)}(0,v) = {\mathbb V}(0,v)$ in addition to the field space derivative being zero one more constraint is obtained: \be e^{i{\beta \over 2} \phi(0)} =0 \label{22} \ee which of course is also automatically obtained from the boundary conditions found previously in the full diagonal case at $e^{-i\xi} = 0$. Note that there is no way to trace the extra constraint (\ref{22}) from Hamiltonian point of view although we have to note that such a constraint is not incompatible with the Hamiltonian.

To conclude we note that in the degenerate case some important information is automatically lost when considering the Hamiltonian description. More precisely, in the degenerate case there is no $\xi$ dependence so constraints of the type (\ref{22}) disappear when examining the boundary conditions from the Hamiltonian viewpoint. Whenever we pass from the most general situation to some special situation some information is lost and inconsistencies between the two descriptions arise. This of course happens only when the $K$-matrix possesses several boundary parameter and some of them are set to zero or to infinity.  We shall examine in the following section a similar situation for the next model of the hierarchy, the $A_2^{(1)}$ theory, and we shall see that the arising inconsistencies may be explained in the same spirit.

\section{The boundary $A_{2}^{(1)}$ case}

We come now to the second member of the hierarchy and the first model
of this class exhibiting both types of distinct boundary
conditions SP and  SNP, i.e. the $A_{2}^{(1)}$ model. In this case we have:
\be \alpha_1 = (1,\ 0), ~~~\alpha_2 = (-{1\over 2},\ {\sqrt 3
\over 2}),~~~~\alpha_3 = (-{1\over 2},\ -{\sqrt 3 \over 2}) \ee
define also the following $3\times 3$ generators \be E_{1} = E_{-1}^t = e_{12}, ~~~~E_2 = E^t_{-2} = e_{23}, ~~~~E_3= E^t_{-3}
= -e_{31}. \ee The diagonal Cartan
generators $H_{1,2}$ are then: \be H_1 ={1\over 2} (e_{11} -e_{22}),
~~~~H_2={1\over 2 \sqrt 3} (e_{11}+e_{22} -2 e_{33}) \ee

Let $T'(x, y, u) = T(x, y, u^{-1})$ and ${\mathbb U}'(x, u) = {\mathbb U}(x, u^{-1})$. Following the logic described previously (see also \cite{ft})
for the sine-Gordon model, we aim at expressing the part associated
to $E_+$, $E_-$ in ${\mathbb U},\ {\mathbb U}'$ respectively
independently of the fields, thus we consider the following gauge
transformation: \be &&
T(x, y, u )= \Omega(x)\ \tilde T(x,y, u)\ \Omega^{-1}(y), \non\\
&& T'(x, y ,u) = \Omega^{-1}(x)\ \tilde T'(x,y,u)\ \Omega(y) \label{gauge} \ee
where we define \be \Omega(x)=\mbox{diag} \Big (\Omega_1(x),\ \Omega_2(x),\ \Omega_3(x) \Big ) = e^{{\beta \over 2} \Phi(x)\cdot H}.\label{gauge0} \ee
From equation (\ref{dif1}) the gauge transformed
operators $\tilde {\mathbb U},\ \tilde{\mathbb U}'$ are expressed as: \be &&
\tilde {\mathbb U}(x,t,u) = \Omega^{-1}(x)\ {\mathbb
U}(x,t,u)\ \Omega(x) - \Omega^{-1}(x)\ {d \Omega(x)\over dx} \non\\
&& \tilde {\mathbb U}'(x,t,u) = \Omega(x)\ {\mathbb U}'(x,t,u)\
\Omega^{-1}(x) - \Omega(x)\ {d \Omega^{-1}(x)\over dx}. \ee After
implementing the gauge transformations $\tilde
{\mathbb U},\ \tilde {\mathbb U}'$ take the following simple forms:
\be \tilde {\mathbb U}(x,t, u) = {\beta \over 2} \mathfrak{F} \cdot H
+{m\over 4} \Big ( u E_+ + {1 \over u} X_- \Big ), ~~~~~ \tilde
{\mathbb U}' (x, t, u) = {\beta \over 2} \hat \mathfrak{F} \cdot H
+{m\over 4} \Big ( u E_- + {1 \over u} X_+ \Big ) \ee where we
define: \be \mathfrak{F} = \Pi -
\partial_x \Phi,~~~\hat \mathfrak{F} = \Pi + \partial_x \Phi, ~~~X_- =e^{- \beta \Phi \cdot H}\ E_-\ e^{\beta  \Phi \cdot H}, ~~~X_+= e^{ \beta \Phi \cdot H}\ E_+\ e^{-\beta  \Phi \cdot H} \ee $\tilde
T,\ \tilde {\mathbb U}$ also satisfy (\ref{dif1}), and $\mathfrak{F},\
\hat \mathfrak{F}$ are vectors with two components $\mathfrak{f}_{i},\ \hat
\mathfrak{f}_{i}, ~~i\in\{1,\ 2\}$ respectively.

Consider again the ansatz (\ref{exp0}) for $\tilde T$, $~\tilde T'$ as
$|u| \to \infty$. As in the previous section inserting
expressions (\ref{expa}) in (\ref{dif1}) one then
identifies the coefficients $W_{ij}^{(k)}$ and $Z_{ii}^{(k)}$.
Indeed from (\ref{dif1}) we obtain the following fundamental
relations: \be && {d Z \over d x} = {\tilde \mathbb U}^{(D)} +
({\tilde \mathbb U}^{(O)}\ W)^{(D)} \non\\
&& {d  W\over dx } + W {\tilde \mathbb U}^{(D)} -{\tilde \mathbb
U}^{(D)}W + W({\tilde \mathbb U}^{(O)}W)^{(D)} -{\tilde \mathbb
U}^{(O)} - ({\tilde \mathbb U}^{(O)}W)^{(O)}=0 \label{form2} \ee
where the superscripts $O,\ D$ denote off-diagonal and diagonal part
respectively. Similar relations may be obtained for $\hat Z,\ \hat
W$, in this case $\tilde {\mathbb U} \to \tilde {\mathbb U}'$. We
omit writing these equations here for brevity.

It will be useful in what follows to introduce some compact notation: \be
{\beta \over 2} \mathfrak{F} \cdot H = \mbox{diag} (a,\ b,\ c),
~~~~~{\beta \over 2} \hat \mathfrak{F} \cdot H = \mbox{diag} (\hat a,\
\hat b,\ \hat c), ~~~~ e^{\beta \alpha_i \cdot \Phi} = \gamma_i.
\label{def} \ee Explicit expressions of $a,\ b,\ c$ and $\gamma_i$
are given by: \be && a = {\beta \over 2}
({\mathfrak{f}_1 \over 2} + {\mathfrak{f}_2 \over 2 \sqrt 3} ), ~~~~b = {\beta
\over 2} (-{\mathfrak{f}_1 \over 2}+ {\mathfrak{f}_2 \over 2 \sqrt 3}),
~~~~c = -{\beta \over 2}  {\mathfrak{f}_2 \over  \sqrt 3}, \non\\
&& \gamma_1 = e^{\beta \phi_1}, ~~~~\gamma_2 = e^{\beta(-{1 \over
2}\phi_1 + {\sqrt 3 \over 2} \phi_2)},
 ~~~~~\gamma_3=e^{\beta(-{1 \over 2}\phi_1 - {\sqrt 3 \over 2} \phi_2)}. \label{def2} \ee
apparently $\hat a,\ \hat b,\ \hat c$ are defined in the same way as $a,\ b,\ c$ but with $\mathfrak{f}_i \to \hat \mathfrak{f}_i$.

The computation of $W,\ \hat W$ is
essential for what follows. First it is important
to discuss the leading contribution of the above quantities as $|u|
\to \infty$. To achieve this we shall need the explicit form of
$Z^{(-1)},\ \hat Z^{(-1)}$: \be Z^{(-1)}(x, y) = {m (x-y)\over 4}
\left( \begin{array}{ccc}
e^{{i \pi \over 3}}     &  & \\
                                          &e^{-{i \pi \over 3}}  &  \\
 &  & -1 \\ \end{array} \right), ~~~\hat Z^{(-1)}(x,y)=  {m (x-y)\over 4}
\left( \begin{array}{ccc}
e^{-{i \pi \over 3}}     &  & \\
                                          &e^{{i \pi \over 3}}  &  \\
 &  & -1 \\ \end{array} \right). \non\\ \label{zz}
\ee
From the formulas (\ref{form2}) the matrices $W^{(k)},\
\hat W^{(k)},\ Z^{(k)},\ \hat Z^{(k)}$ may be determined. In
particular, we write below explicit expressions of these matrices
for the first orders, which will be necessary in the subsequent sections (see also \cite{doikouatft}):
\be && W^{(0)} = \hat W^{(0)}= \left(
\begin{array}{ccc}
0                    & e^{{i \pi \over 3}}  & 1 \\
e^{{i \pi \over 3}}  & 0                    & -1 \\
e^{{2i \pi \over 3}} & e^{-{i \pi \over 3}} & 0\\ \end{array} \right), \non\\
&& {m\over 4} W^{(1)}= \left( \begin{array}{ccc}
0       & e^{{2i \pi \over 3}} a  & c \\
-a      & 0                       & b \\
e^{{i \pi \over 3}}c  & -b        & 0\\ \end{array} \right),
~~~~{m\over 4} \hat W^{(1)} = \left( \begin{array}{ccc}
0       & -\hat b  & -\hat a \\
-e^{-{i \pi \over 3}}\hat b      & 0                       & -\hat c \\
\hat a  & -e^{{i \pi \over 3}}\hat c        & 0\\ \end{array}
\right). \ee
For computing the boundary conserved quantities, energy and momentum, we shall in addition need the following expressions:
\be
&& {d Z_{11}^{(1)}\over dx} = {e^{-{i\pi \over 3}}\over 3} {m
\over 4}(\gamma_1 +\gamma_2 + \gamma_3)  +{4 e^{-{i\pi \over
3}}\over 3m} (a' -c') +{4e^{-{i\pi \over 3}} \over 6m} (a^2
+b^2 +c^2) \non\\ && {d Z_{22}^{(1)} \over dx} = {e^{{i\pi \over
3}} \over 3} {m \over 4}(\gamma_1 +\gamma_2 + \gamma_3)  +{4
e^{{i\pi \over 3}} \over 3m} (b' -a') +{4 e^{{i\pi \over
3}}\over 6m} (a^2 +b^2 +c^2) \non\\
&& {d \hat Z_{11}^{(1)} \over dx} = {e^{{i\pi \over 3}} \over 3}
{m \over 4}(\gamma_1 +\gamma_2 + \gamma_3) -{4 e^{{i\pi \over
3}} \over 3m} (\hat b' -\hat a') +{4 e^{{i\pi \over 3}}\over 6m}
(\hat a^2 +\hat b^2 +\hat c^2)
\non\\
&& {d \hat Z_{22}^{(1)} \over dx} = {e^{-{i\pi \over 3}}\over 3}
{m \over 4}(\gamma_1 +\gamma_2 + \gamma_3)  +{4 e^{-{i\pi
\over 3}}\over 3m} (\hat b' - \hat c') +{4 e^{-{i\pi \over 3}}
\over 6m} (\hat a^2 +\hat b^2 +\hat c^2). \ee

\subsection{SNP boundary conditions}

We shall focus in this section on the analysis of the SNP integrable boundary conditions in
$A_2^{(1)}$ ATFT.
Comparison with some already known results \cite{durham} will
validate our approach, which then presents the advantage of being systematically
implementable once a non dynamical ``boundary'' matrix
is chosen. The boundary ${\mathbb V}$-operator in this case is
given by (\ref{final2}), (\ref{boundary2}). We assume here for simplicity, but
without losing generality, Schwartz boundary conditions at $x =-L$ and $K^-
\propto {\mathbb I}$ (see also \cite{doikouatft}). Taking into account
the ansatz for the monodromy matrix (\ref{exp0}) as well as bearing in
mind that as $u\to \infty$ the main contribution for the diagonal
terms comes from $e^{Z_{33}},\ e^{\hat Z_{33}}$ (see also
\cite{doikouatft}), we conclude \be {\mathbb V}(x,t,u,v) &=&
\Big [ (1+W(x,u))^{-1} \Omega^{-1}(x)
r(uv^{-1}) \Omega(x) (1+W(x))\Big ]_{33} \non\\ & + & \Big [
(1+\hat W^t(x,u)) \Omega^{-1}(x) r^{t_1}(u^{-1}v^{-1})
\Omega(x)((1+\hat W(x))^{-1})^t \Big ]_{33}. \label{vv}\ee We recall
that the superscript $^{t_1}$ denotes transposition in the first
space. Also, in the expressions with `hat' we simply consider
$\Phi \to- \Phi$. For further technical details we refer the interested reader to
\cite{doikouatft}. Note that in this case the limit $u \to \infty$ is easier to consider due to the expressions (\ref{zz}). In any case, although technically more involved, one can show that the $u\to -\infty$ limit provides the same conserved quantities and Lax pairs. In the following we shall expand expression (\ref{vv}), so
we need expansions of all the involved quantities: \be r(uv^{-1})
 \sim r^{(0)} + {u^{-1}}r^{(1)} + {\cal O}(u^{-2}),
~~~~~r^{t_1}(u^{-1}v^{-1})  \sim -r^{(0)} - u^{-1} \hat  r^{(1)} +
{\cal O}(u^{-2}) \ee where we define: \be r^{(0)} &=& \sum_{i=1}^3
e_{ii}, ~~~~r^{(1)}= 2 v (e_{21}\otimes e_{12} + e_{32}\otimes
e_{23} + e_{13}\otimes e_{31}) \non\\ \hat r^{(1)} & = & 2v^{-1}
(e_{21} \otimes e_{21} + e_{32} \otimes e_{32} +e_{13}\otimes
e_{13}) \label{r0} \ee
From the first order of the expansion of the
${\mathbb V}$-operator and
after multiplying with a factor of $-{3m \over 8}$ we obtain: \be
{\mathbb V}(x,t,v)= -{\beta \over 2} \Phi'(x,t)\cdot H +{m v \over 4}
\Omega(x,t) E_+\Omega^{-1}(x,t)-{mv^{-1}\over 4} \Omega^{-1}(x,t) E_-
\Omega(x,t) \label{vbulk0}\ee which as anticipated coincides with the bulk ${\mathbb V}$-operator of $A_2^{(1)}$ (\ref{lpair}).

In order to obtain the explicit form of the boundary ${\mathbb V}$-operator
we should also review known results on the
solutions of the reflection equation for SNP boundary conditions.
The generic solution for the $A_{n}^{(1)}$ case in the principal
gradation are given by \cite{gand, ann2}: \be && K(\lambda) = (g
e^{\lambda} + \bar g e^{-\lambda}) \sum_{i=1}^{n+1} e_{ii} +
\sum_{i>j} f_{ij} e^{ \lambda - {2\lambda \over n+1} (i-j)}
e_{ij}+ \sum_{i<j} f_{ij} e^{ -\lambda - {2\lambda \over n+1}
(i-j)} e_{ij} \non\\ && g =q^{-{1\over 2} + {n+1 \over 4}}
~~~~\bar g = \pm q^{{1\over 2} - {n+1 \over 4}}, ~~~f_{ij} = \pm
q^{-{n+1 \over 4}}, ~~~~f_{ji} = q^{{n+1 \over 4}}, ~~~i < j.
\label{kp} \ee In order to effectively compare with the results of
\cite{gand} as well as being compatible with \cite{durham} we
always express both $r$ and $K$ matrices in the
principal gradation (see also \cite{doikouatft}). The parameter $q$ in the solution (\ref{kp})
is the parameter of the underlying quantum algebra (quantum case) $U_q(\widehat{sl_3})$.
It is clear that we
are dealing here with the classical
limit $q \to 1$ of the later solution (\ref{kp})
compatible with the classical quadratic algebra (\ref{refc2}). In this limit:
$g\to 1,\ \bar g \to \pm 1,\ f_{ij} \to \pm 1,\ f_{ji} \to 1$, $i<j$.

We come back now to the $A_2^{(1)}$ case. Recall that $K^+(u) = K^t(u^{-1})$ then
$K^+$ is a $3 \times 3$ matrix written explicitly as: \be &&
K^+(u) = u^{{3\over 2}} \bar G + u^{{1\over 2}} \bar F +
u^{-{1\over 2}} F + u^{-{3\over 2}} G ~~~~~\mbox{where} \non\\
&& G = g\ {\mathbb I}, ~~~\bar G = \bar g\ {\mathbb I}, \non\\
&& \bar F = f_{12}\ e_{21} + f_{23}\ e_{32} + f_{31}\ e_{13}, \non\\
&& F = f_{21}\ e_{12} +f_{32}\ e_{23}  +f_{13}\ e_{31} \label{k1}
\ee and the coefficients $g,\ \bar g,\ f_{ij}$ are given in
(\ref{kp}) with $n=2$ and $q \to 1$. Note, as in the previous case for the
sine-Gordon model, that the boundary case for $x=0$ has to be
treated separately. Indeed, in this case the operator ${\mathbb
V}(0,u)$ takes the form: \be\tilde {\mathbb V}(x=0) &=& \Big [
(1+\hat W^t(0))
\Omega^{-1}(0) K^+(u) \Omega (1+W(0)) \Big ]_{33} \non\\
&\times& \Big \{ \Big [ (1+\hat W^+(0)) \Omega^{-1}(0) K^+(u)
r(uv^{-1})\Omega(0) (1+W(0)) \Big ]_{33} \non\\ &+& \Big [ (1+\hat
W^t(0)) \Omega^{-1}(0) r^{t_1}(u^{-1}v^{-1}) K^+(u) \Omega(0)
(1+W(0)) \Big ]_{33} \Big \} \ee again using the standard
procedure we obtain from the
first order of the expansion and after multiplying with an
overall factor $-{3m \over 8}$: \be {\mathbb V}^{(b)}(0,t,u) &=&
{\mathbb V}(0,t,u) + \Delta {\mathbb V}(0,t,u), ~~~~\mbox{where} \non\\
\Delta {\mathbb V} (0,t,u) &=& {\beta \over 2} \Phi'(0) \cdot H
+{m\over 8\bar g }\Big ( (\Omega_1(0) \Omega_2^{-1}(0) f_{12} + \Omega_1^{-1}(0)
\Omega_3(0)f_{31} )e_{11} \non\\ &+& (\Omega_2(0)
\Omega_3^{-1}(0)f_{23} -\Omega_1(0) \Omega_2^{-1}(0)f_{12} )e_{22}
\non\\ &-&  (\Omega_2(0) \Omega_3^{-1}(0)f_{23} +
\Omega_1^{-1}(0)\Omega_3(0)f_{31})e_{33} \Big )\non\\ \ee ${\mathbb V}$ is the bulk
${\mathbb V}$-operator given in (\ref{vbulk0}).
The equations of motion are again obtained from the `bulk' zero curvature condition:
\be -\ddot{\Phi}(x,t)+ \Phi''(x,t) ={m^2 \over 2 \beta} \sum_{i=1}^3 \alpha_i e^{\beta \alpha_i \cdot \Phi(x,t)}. \label{lbulk} \ee
Using the same argument as in the previous section we conclude: \be \Delta {\mathbb V}(0) & = & 0 \non\\
\Rightarrow  \phi_1'(0) &=& -{m \over 2\bar g \beta } \Big (2 f_{12} e^{{\beta \over 2}
\alpha_1 \cdot \Phi(0)} -f_{23} e^{{\beta \over 2}
\alpha_2 \cdot \Phi(0)}+f_{31} e^{{\beta \over 2}
\alpha_3 \cdot \Phi(0)}\Big )\non\\  \phi_2'(0) &=& -{m \sqrt 3 \over 2\bar g \beta } \Big (f_{23} e^{{\beta \over 2}
\alpha_2 \cdot \Phi(0)}+f_{31} e^{{\beta \over 2}
\alpha_3 \cdot \Phi(0)}\Big )\ee
The latter boundary conditions agree as expected with those analyzed in \cite{durham}.

The quantities found above are associated to the first non-trivial
boundary integral of motion: \be {\cal H}_1^{(b)} =\int_{-L}^{0} dx \Big( \sum_{i=1}^2 (\pi_i^2 +
\phi_i^{'2}) + {m^2 \over \beta^2} \sum_{i=1}^3 e^{\beta \alpha_i
\cdot \Phi}  \Big ) +{2 m \over \bar g \beta^2} \Big ( f_{12}
e^{{\beta \over 2} \alpha_1 \cdot \Phi(0)} +f_{23} e^{{\beta \over
2} \alpha_2 \cdot \Phi(0)} -f_{31}
e^{{\beta \over 2} \alpha_3 \cdot \Phi(0)}   \Big ). \non\\
\label{bham} \ee It is quite easy to check the consistency of the Lax formulation procedure in \cite{avandoikou}, i.e. that the Hamiltonian above
leads exactly to the same equations of motion and boundary conditions as the
zero curvature conditions. More precisely, the equations of motion from
the Hamiltonian are obtained through: \be && {\partial \phi_i(x ,t)
\over \partial t} = \{{\cal H}, ~\phi_i(x, t)  \}, ~~~~~
{\partial \pi_i(x ,t) \over \partial t} = \{{\cal H}, ~
\pi_i(x, t)  \} \non\\ && x \in [-L,~ 0] \label{eqmo}\ee

The boundary Lax pair in the SNP case was also constructed in \cite{durham}. The boundary conditions and the Hamiltonian extracted in \cite{doikouatft} coincide with the ones of \cite{durham}; comparison between our Lax pair and that derived in \cite{durham} shows that the relevant Lax pairs are also the same. Although we have to note that in \cite{durham} everything is expressed in terms of $\theta$ and $\delta$ functions given that the Lax pair is derived taking into account some overlapping boundary regime, which in the present framework is unnecessary. Moreover in \cite{durham} the Lax pair construction requires a priori knowledge of the Lagrangian, given that the boundary Lax pair explicitly contains the Lagrangian boundary contribution. In our formulation on the other hand the Lax pair is derived from first principles from the expansion of (\ref{final2}), (\ref{boundary2}), and requires no a priori knowledge of the boundary terms in the associated Hamiltonian, although the relevant Hamiltonian may be simultaneously obtained from the expansion of $\ln t(\lambda)$.

This ends our discussion on the sine-Gordon and SNP $A_2^{(1)}$ ATFT.
We have been able to systematically derive consistent explicit expressions
from first principle derivation for their Lax equations by evaluating
{\it exactly} the boundary contribution.
Comparison with the case-by-case derivation in \cite{durham} shows the same
results. Our method is therefore validated and can be now extended to a much broader set of models beyond the ATFT. The next example will be the yet untreated case of $A_2^{(1)}$ ATFT under the ``soliton-preserving'' reflection algebra (see also \cite{doikouatft} on SP boundary conditions within ATFT).

\subsection{SP boundary conditions}

We shall now examine the SP boundary conditions (see \cite{doikouatft}).
We shall consider below two cases with the $K$-matrix being non-diagonal and diagonal respectively.

\subsubsection{Non-diagonal $K$-matrix}

We choose for $K$ the non-diagonal solution of
the reflection equation found in \cite{kumu, nich}, and demonstrate how this particular choice
of $K$-matrix contributes to the integrals of motion, and the relevant boundary Lax pair. We consider for simplicity Schwartz boundary conditions at the end point $x =-L$,
whereas the right boundary is described by this $K$-matrix \cite{kumu, nich}
\be && K^{+}(u)= {\cal V}\ K(u)\ {\cal V} ~~~~\mbox{with} ~~~~~{\cal V}= \mbox{diag}(1,\ e^{2\lambda \over 3},\ e^{4\lambda \over 3})\non\\ && K(u)= (e^{4 \lambda} -1){\mathrm g}^2 +(e^{4 \lambda} -1)(\xi e^{2\lambda} - \alpha-\beta){\mathrm g} - (\xi^2 e^{4 \lambda} - (\alpha + \beta) \xi e^{2 \lambda} + \alpha \beta ) {\mathbb I} \label{kmatrixsp} \ee where we define the $3 \times 3$ matrix \cite{kumu, nich}: \be {\mathrm g} = (\alpha +\beta ) e_{11} + x^+ e_{12} + x^- e_{21}, ~~~~~x^+ x^- = - \alpha \beta \label{bb} \ee $\alpha,\ \beta,\ \xi$ are free independent boundary parameters, and $x^{\pm}$ satisfy (\ref{bb}).
To proceed with the expansion of $\ln t(\lambda)$ and  ${\mathbb V}$ in powers of
$u^{-1}$ we shall need the expansions of $r,\ \hat r$, $K^{+}$, (as $|u| \to \infty $): \be
\hat r(uv) & \sim & r^{(0)} + u^{-1} \hat r^{(1)} + {\cal
O}(u^{-1}) \non\\ K^+(u) & \sim & K^{(0)} + u^{-1} K^{(1)} +{\cal O}(u^{-2}). \ee
where $r^{(0)}$ is defined in (\ref{r0}) and \be && \hat r^{(1)} = 2v^{-1} (e_{12}
\otimes e_{21} + e_{23} \otimes e_{32}+ e_{31}\otimes e_{13} ),\non\\ && K^{(0)} = -\xi^2 e_{33} +\xi x^+ e_{12} + \xi x^- e_{21}, ~~~~K^{(1)} =  \xi (\alpha +\beta) e_{11} \ee
We shall consider henceforth for simplicity $x^+= x^-$, and also set ${\mathrm y} = x^{\pm} \xi$.
The integrals of motion follow from the asymptotic expansion
 of the logarithm of the open transfer matrix as $i u \to \pm \infty$, i.e.
\be && \ln t(i u \to \infty) = \sum {{\mathbb I}_n \over u^n} = \sum {Z_{11}^{(n)} - \hat Z_{11}^{(1)} \over u^n} +\sum {{\mathrm h}_n \over u^n} \non\\ && \ln t(i u \to -\infty) = \sum {\tilde {\mathbb I}_n \over u^n} = \sum {Z_{22}^{(n)} - \hat Z_{22}^{(1)} \over u^n} +\sum {\tilde  {\mathrm h}_n \over u^n}\ee More technical details on such expansions will be found by the interested reader in \cite{doikouatft}.
The integrals of motion obtained from the first order of the expansion
are given by: \be {\mathbb I}_1 &=& -{\beta^2 \over 12 m} ( {\cal P}^{(b)} + i \sqrt 3{\cal H}^{(b)}) , \non\\ \tilde {\mathbb I}_1 &=& -{\beta^2 \over 12m} ( {\cal P}^{(b)} - i \sqrt 3 {\cal H}^{(b)}). \label{zz21} \ee The momentum and energy are directly obtained from
the above conserved quantities and defined as: \be && {\cal P}^{(b)} = \int_{-L}^0 dx \sum_{i=1}^2
\Big (\pi_i\ \phi_i' - \pi_i'\ \phi_i \Big ) + \sum_{i=1}^2
\pi_i(0)\ \phi_i(0) - {12 m \over \beta^2 \Delta} \xi (\alpha +\beta) \Omega_1^2(0) \non\\ & & - {8 \over \beta^2 \Delta } \Big \{ {\mathrm y} \Omega_1(0) \Omega_2(0) \Big(-4a(0) +b(0) -4\hat a(0) + \hat b(0) \Big ) - \xi^2 \Omega_3^2(0) \Big (c(0) -b(0)+ \hat c(0) -\hat b(0) \Big )  \Big \}
\non\\
&& {\cal H}^{(b)} = \int_{-L}^0 dx \Big (
\sum_{i=1}^2 (\pi_i^2 + \phi_i^{'2} ) + {m^2 \over \beta^2}
\sum_{i=1}^3 e ^{\beta \alpha_i \cdot \Phi} \Big ) \non\\ &&- {8 \over
\beta^2 \Delta } \Big \{ {\mathrm y} \Omega_1(0) \Omega_2(0) \Big(c(0) - a(0) - \hat c(0) + \hat a(0) \Big) +\xi^2 \Omega_3^2\Big ( c(0)-b(0) -\hat c(0) + \hat b(0) \Big )\Big \}. \non\\ \label{ph1} \ee where $\Delta = {\mathrm y} \Omega_1(0) \Omega_2(0) - \xi^2 \Omega_3^2(0)$, and $\Omega$ and $a,\ b,\ c,\ \hat a,\ \hat b,\ \hat c$ have been previously defined.
As already pointed out in \cite{doikouatft} the two boundary cases exhibit
essential differences: in SNP the $c$-number $K$-matrix contains no free
parameters, and consequently no such parameters occur in the
deduced integrals of motion. In the SP case on the other hand the
$K$-matrix contains free parameters, which explicitly appear in
the boundary integrals of motion. Moreover, in the SNP case, only the boundary
Hamiltonian belongs to the family of commuting quantities, whereas now both
boundary Hamiltonian and momentum turn out to be conserved quantities.
Note that we are here in a situation where the terminology ``boundary effects'' is misleading, since
it suggests that no conserved
momentum could exist due to the presence of a physical boundary. As we have shown here this is not true,
indicating that the physical interpretation of this field theory is more subtle and has to be further
explored.

We shall now derive the associated boundary ${\mathbb V}$-operator.
From the two different limits ($i u\to \pm \infty$) we obtain essentially two quantities
from the first order of each expansion: \be {\mathbb
V}_1^{(1)}(x,t,v) &=& {4  \over 3 m}\Big ( {\mathbb U}(x,t,v)
-i\sqrt 3 {\mathbb V}(x,t,v) \Big ) ~~~~\mbox{from} ~~~~i u \to
\infty \non\\ {\mathbb V}_2^{(1)}(x,t,v) &=& {4  \over 3m
}\Big ( {\mathbb U}(x,t,v) + i\sqrt 3 {\mathbb V}(x,t,v) \Big )
~~~~\mbox{from} ~~~~i u \to -\infty \label{VV} \ee where ${\mathbb
V},{\mathbb U} $ is the Lax pair of
the $A_2^{(1)}$ theory defined in (\ref{lpair}). The bulk operator as already mentioned is independent of the choice of $K$-matrix.

The expressions of the ${\mathbb V}$-operator for each
end point are given below: \be
{\mathbb V}(0,t,u,v) &=& \Big [(1+\hat W(0))^{-1} \Omega(0) K^+(u)\Omega(0) (1+W(0)) \Big]^{-1}_{jj} \non\\
& \times & \Big \{ \Big [ (1+\hat W(0))^{-1} \Omega(0)K^+(u)
r(uv^{-1}) \Omega(0)(1+W(0)) \Big ]_{jj} \non\\ &+& \Big [(1+\hat
W(0) )^{-1} \Omega(0) \hat r(uv)K^+(u) \Omega(0)(1+W(0))
\Big ]_{jj} \Big \},~~~~j \in \{1,\ 2\}.\non\\ \ee
At the boundary point we get:
\be  {\mathbb V}_1^{(b)}(0, v) &=& {2\over m} \Big ({\mathbb U}^{(b)}(0, v) - i\sqrt 3 {\mathbb V}^{(b)}(0, v) \Big ), ~~~~i u \to \infty \non\\  {\mathbb V}_2^{(b)}(0, v) &=& {2\over m} \Big ( {\mathbb U}^{(b)}(0, v)  + i\sqrt 3 {\mathbb V}^{(b)}(0, v) \Big ), ~~~~i u \to -\infty. \label{Vb} \ee
Let us focus on ${\mathbb V}^{(b)}$ (\ref{Vb}), which is associated
to the Hamiltonian of the system (\ref{ph1}) as will become transparent in the following. Note that ${\mathbb U}^{(b)}$ similarly plays the role of the ${\mathbb V}$-operator associated to the momentum of the system.
We then define: \be {\mathbb V}^{(b)}(0, v) &=& {1 \over \Delta^2} \xi^2 \Omega_3^2(0) {\mathrm y} \Omega_1(0) \Omega_2(0) \Big (c(0) -\hat c(0)\Big ) \Big (e_{11} +e_{22} -2 e_{33}\Big ) \non\\ & +& {m v \over 2 \Delta} \Big (-\xi^2 \Omega_2(0) \Omega_3(0) e_{23} - {\mathrm y} \Omega_2(0) \Omega_3(0)  e_{31}\Big ) \non\\ &-& {m v^{-1} \over 2 \Delta} \Big (-\xi^2 \Omega_2(0) \Omega_3(0) e_{32} - {\mathrm y} \Omega_2(0) \Omega_3(0)  e_{13}\Big ) \non\\ {\mathbb U}^{(b)}(0,v) &=& {1\over \Delta^2} \xi^2 \Omega_3^2(0) {\mathrm y} \Omega_1(0) \Omega_2(0) \Big (a(0) -b(0)  +\hat a(0)- \hat b(0)\Big )\Big (e_{11} +e_{22} -2 e_{33}\Big ) \non\\ &+& {\Omega_1^2(0) m \xi \over 2 \Delta}\Big (\alpha(0) +\beta(0)\Big ) \Big ({\mathrm y} \Omega_1(0)\Omega_2(0)(e_{11}-e_{22})-2\xi^2\Omega_3^2(0) (e_{11} - e_{33}) \Big ) \non\\ &+& {m v \over 2 \Delta}  \Big (2 {\mathrm y} \Omega_1^2(0) e_{12}-\xi^2 \Omega_2(0) \Omega_3(0) e_{23}  + {\mathrm y} \Omega_2(0) \Omega_3(0)  e_{31} \Big ) \non\\ &+& {m v^{-1} \over 2 \Delta} \Big (2 {\mathrm y} \Omega_1^2(0) e_{21}-\xi^2 \Omega_2(0) \Omega_3(0) e_{32}  + {\mathrm y} \Omega_2(0) \Omega_3(0)  e_{13} \Big ). \non \ee
We shall focus on the equations of motion emerging from the Hamiltonian ${\cal H}^{(b)}$ via (\ref{eqmo}). We expect that identical equations arise from the zero curvature condition. Indeed the bulk part gives rise to the equations of motion that coincide with the familiar ones (see equations (\ref{lbulk})).

From the Hamiltonian derivation the boundary conditions arise by requiring the boundary terms (proportional to $\phi'_i(0)$) to vanish, yielding: \be \Phi'(0) =0  ~~~~~{\mathrm y} \Omega_1(0) \Omega_2(0)= -\xi^2 \Omega_3^2(0). \label{further1} \ee
From the zero curvature condition however, as analyzed in section 2, we require ${\mathbb V}(0) = {\mathbb V}^{(b)}(0)$, and we end up with an extra constraint in addition to the ones (\ref{further1}) emerging from the Hamiltonian, i.e. \be \Omega_1(0) \Omega_2^{-1}(0) =0. \ee Although this extra constraint is compatible with the Hamiltonian description it is nevertheless missing when analyzing the boundary conditions from the Hamiltonian point of view. This `missing' information may be associated to the fact that the $K$-matrix we have chosen (\ref{kmatrixsp}) is not the most general matrix to start with, and exhibit some degenerate or at least non-generic behavior, as we shall comment in the final section.
Finally one may derive the equations of motion from the boundary ${\mathbb V}$-operator associated to the momentum ${\cal P}$ along the lines described above.

To conclude we have here also been able to explicitly derive exact expressions of the boundary Hamiltonian and momentum --both conserved quantities for the particular boundary conditions-- as well as the associated boundary Lax pairs.

\subsubsection{Diagonal $K$-matrix}

We shall now  focus on diagonal degenerate solutions of
the reflection equation \cite{dvg} given by the following
expressions (in the principal gradation): \be K_{(l)}(\lambda,\
\xi) = \sinh (\lambda + i\xi) e ^{-\lambda} \sum_{j=1}^l
e^{-{4\lambda \over n+1}(j-1)}e_{jj} + \sinh (-\lambda + i \xi)e
^{\lambda} \sum_{j=l+1}^n e^{-{4\lambda \over n+1}(j-1)}e_{jj} \ee
(recall $u = e^{2\lambda \over n+1}$ ). As we have seen from the previous section on the sine-Gordon degenerate $K$-matrices give rise to inconsistencies. Some information is usually lost from the Hamiltonian point view when taking the limit from the generic non-degenerate $K$ matrix to the degenerate one, which is a special case of the generic solution. To obtain the $K$-matrix
in the homogeneous gradation one implements a simple gauge transformation (see e.g. \cite{doikouatft}).

In the $A_2^{(1)}$ case we end up with two types of diagonal boundary matrices
corresponding to the two possible values $l=1,\ 2$. We shall
consider an example here to demonstrate how the particular choice
of boundary $K$-matrix contributes to the integrals of motion.
It is practical for the following to consider a non-trivial left boundary
described by $K_{(1)}$, and a right boundary described by the
$K_{(2)}$-matrix, i.e. \be K^{+}(u,\ \xi^+)= K_{(1)}(u^{-1},\
\xi^+), ~~~~~~K^-(u,\ \xi^-) =K_{(2)}(u,\ \xi^-). \ee
We now proceed with the expansion of ${\mathbb V}$ in powers of
$u^{-1}$. The bulk part,  for $x \neq \ 0,\ -L$, is given in (\ref{VV}).

To obtain the generic results with the least effort it is practical to
consider the two different types of $K$-matrices to each end of
the theory. The expressions of the ${\mathbb V}$-operator for each
end point $x_b =0,\ -L$ are given by \be {\mathbb
V}_1^{(b)}(x_b,t,v) &=& {4  \over m}\Big ( {\mathbb U}^{(b)}(x_b,t,v)
-i\sqrt 3 {\mathbb V}^{(b)}(x_b,t,v) \Big ) ~~~~\mbox{for} ~~~~i u \to
\infty \non\\ {\mathbb V}_2^{(b)}(x_b,t,v) &=& {4  \over m
}\Big ( {\mathbb U}^{(b)}(x_b,t,v) + i\sqrt 3 {\mathbb V}^{(b)}(x_b,t,v) \Big )
~~~~\mbox{for} ~~~~i u \to -\infty.  \ee
We are mostly interested in ${\mathbb V}^{(b)}$, which we shall associate to the Hamiltonian of the system as will become transparent in the following. Note that ${\mathbb U}^{(b)}$ plays the role of the ${\mathbb V}$-operator associated to the momentum of the system.
We then have \be {\mathbb V}^{(b)}(0,v) &=&  { m  \over 4} \Omega_2^2(0) \Omega_3^{-2}(0) \Big (v e_{23} -v^{-1}  e_{32} \Big ) \non\\ {\mathbb V}^{(b)}(-L,v) &=&  {m  \over 4}  \Omega_1^2(-L) \Omega_2^{-2}(-L) \Big (v e_{12} + v^{-1} e_{21} \Big ). \ \ee
Just for the record we also give: \be {\mathbb U}^{(b)}(0,v) &=& {m\over 2} e^{-2 i \xi^+}  \Omega_1^2(0) \Omega_3^{-2}(0)\Big( e_{33} - e_{11} \Big) + { m  \over 4} \Omega_2^2(0) \Omega_3^{-2}(0) \Big (v e_{23} +  v^{-1}  e_{32} \Big )\non\\
{\mathbb U}^{(b)}(-L,v) &=& {m\over 2} e^{-2 i \xi^+} \Omega_1^2(-L) \Omega_3^{-2}(-L)  \Big ( e_{33} - e_{11} \Big )+ {m  \over 4}  \Omega_1^2(-L) \Omega_2^{-2}(-L) \Big (v e_{12} + v^{-1} e_{21} \Big ). \non \ee
The integrals of motion emerging from the first order
of the asymptotics of the transfer matrix as $i u \to \pm \infty$
are given by: \be {\mathbb I}_1 &= & = -{\beta^2 \over 12 m} ( {\cal P}^{(b)} + i \sqrt 3{\cal H}^{(b)}) , \non\\ \tilde {\mathbb I}_1 &=& -{\beta^2 \over 12m} ( {\cal P}^{(b)} - i \sqrt 3 {\cal H}^{(b)}). \label{zz2} \ee The momentum and energy are directly obtained from
the above conserved quantities and defined as: \be {\cal P}^{(b)} &=& \int_{-L}^0 dx \sum_{i=1}^2
\Big (\pi_i\ \phi_i' - \pi_i'\ \phi_i \Big ) + \sum_{i=1}^2
\pi_i(0)\ \phi_i(0) + {8 \over \beta} \alpha_2 \cdot \Pi(0)  + {12
m \over \beta^2} e^{-2i \xi^+} \Omega_1^2(0) \Omega_3^{-2}(0)
\non\\ & -& \sum_{i=1}^2 \pi_i(-L)\ \phi_i(-L) - {8 \over \beta}
\alpha_1 \cdot \Pi(-L) + {12 m
\over \beta^2} e^{-2i \xi^-} \Omega_1^2(-L) \Omega_3^{-2}(-L) \non\\
{\cal H}^{(b)} &=& \int_{-L}^0 dx \Big (
\sum_{i=1}^2 (\pi_i^2 + \phi_i^{'2} ) + {m^2 \over \beta^2}
\sum_{i=1}^3 e ^{\beta \alpha_i \cdot \Phi} \Big ) + {8 \over
\beta } \alpha_2 \cdot \Phi'(0)  - {8 \over \beta } \alpha_1 \cdot
\Phi'(-L). \label{ph} \ee Naturally the two boundary cases are
distinguished; in SNP the $c$-number $K$-matrix contains no free
parameters, and consequently no free parameters occur in the
entailed integrals of motion. In the SP case on the other hand the
$K$-matrix contains free parameters, which explicitly appear in
the boundary integrals of motion.

As mentioned earlier upon the `boundary' quantity ${\mathbb V}^{(b)}$ is associated to the Hamiltonian ${\cal H}^{(b)}$. We shall focus on the equations of motion emerging from the Hamiltonian via (\ref{eqmo}). The same equations should emerge from the zero curvature condition. Indeed the bulk part gives rise to the equations of motion that coincide with the familiar ones. Note however that from the Hamiltonian the boundary condition are: \be \Phi'(0) = \Phi'(-L) =0 \label{bc3} \ee whereas from the condition $ {\mathbb V}^{(b)}(x_b)={\mathbb V}(x_b) $ in addition to (\ref{bc3}) extra constraints involving the boundary field appear, more precisely:
\be \Omega_2^{-2}(0)\ \Omega_3^{2}(0) \to 0 ~~~~~~~\Omega_1^{-2}(-L)\ \Omega_2^{2}(-L)\to 0. \label{further} \ee Again as in the non-diagonal case a discrepancy between the two descriptions is apparent due to the degenerate nature of the $K$-matrix.

\section{Comments}

We have analyzed via the boundary Lax pair formulation two distinct types of boundary conditions. The SP case presents a particular interest given that certain subtle technical points arise requiring further clarification. Let us briefly comment on the difficulties emerging in this context.

We see that when the $K$-matrix possesses several free boundary parameters,
discrepancies between the Hamiltonian and Lax description, which however are expected to
be equivalent, are seen to emerge when some parameters are equal to zero
(e.g. for SP boundary conditions in $A_2^{(1)}$ case and even
in sine-Gordon). This is an intriguing point and its resolution is probably associated to
defining a notion of ``appropriate choice'' of the $c$-number representation ($K$-matrix)
of the reflection equation. A generic non-diagonal solution of the reflection
equation is required as a starting point. For instance in
sine-Gordon model, amongst  all non-diagonal solutions (\ref{kmatrix0}) the only ``legitimate''
solution is the one with $x^+ = - x^-$. Such a restriction is here dictated
by the requirement of a consistent asymptotic behavior of the generating function.

It is worth noting that diagonal $K$-matrices we have considered here as well as non-diagonal ones of the type \cite{abadrios} in $A_2^{(1)}$ (SP) have only two
distinct eigenvalues (homogeneous gradation) hence their spectra are
doubly degenerate. Similar discrepancies occur as we have seen in the context of sine-Gordon model when choosing to consider $K \propto {\mathbb I}$ (homogeneous
gradation). Such a matrix has just one distinct eigenvalue so it is
doubly degenerate, leading again to inconsistencies. In the example we considered here in the $A_2^{(1)}$ case even though the non-diagonal matrix $K$ matrix is not degenerate (homogeneous gradation), the matrix ${\mathrm g}$ in (\ref{kmatrixsp}) has a zero eigenvalue, and presumably this is the point that creates the problem in this case.

In a more algebraic framework, all solutions of the reflection equation --the solutions of the quantum reflection equation satisfy the classical reflection equation as well-- considered in the present work are representations of the cyclotomic Hecke algebra (see e.g. \cite{doma, kumu, nich}). The generators $g_l,\ g_0$, $l =1, \ldots, N-1$ of the cyclotomic algebra $C_N^{(n)}$ satisfy the following set of constraints: \be && g_{l}\ g_{l+1}\ g_{l} = g_{l+1}\ g_{l}\ g_{l+1}, ~~~~~g_{1}\ g_{0}\ g_{1}\ g_{0} = g_{0}\ g_{1}\ g_{0}\ g_{1},  \non\\
&& (g_l - q)(g_l + q^{-1}) =0, ~~~~~\prod_{\alpha=1}^n(g_{0}-\xi_{\alpha})=0 \non\\  && [ g_{l},\ g_{m}]  =0, ~~~|l-m| >1, ~~~~~[g_{0},\ g_{l}] =0, ~~l > 1
\label{braid} \ee  One expects that the most generic solution for the $A_{n-1}^{(1)}$ case should be expressed in terms of representations of the generator $g_0$ with all $\xi_{\alpha} \neq 0$ and $\xi_{\alpha} \neq \xi_{\beta}\ \forall\ l \neq k$, i.e. (see also (\ref{kmatrixsp}))
\be K(\lambda) = \sum_{\alpha=0}^{n-1} c_{\alpha}(\lambda)\ g_0^{\alpha} \ee
the general solution should be thus dictated by the rank of the algebra. Note in particular that for the sine-Gordon model $n=2$ one recovers the boundary Temperley-Lieb algebra (see e.g. \cite{doma}).
In this context all solutions that give rise to inconsistencies are special in the sense that are either degenerate $\xi_{\alpha} = \xi_{\beta}$ or correspond to a case with at least one zero eigenvalue $\xi_{\alpha}=0$.

To summarize:
special cases of $K$-matrices give rise to inconsistencies,
hence one needs to consider the most general possible solutions of the
reflection equation with distinct independent boundary parameters.
In the SNP case no extra free boundary
parameters appear and no extra constraints among
the boundary fields occur. In the $A_2^{(1)}$ SP case we conjecture that any
generic (non-degenerate) non-diagonal solution with
free boundary parameters will
be appropriate. For the moment we have no such a generic matrix at our disposal, but the inconsistencies arising give us a strong hint that there should exist
solutions with more boundary parameters. We thus conjecture
that the $K$ matrix (\ref{kmatrixsp}) will turn out to follow from some
yet-to-be found general solution via a limit process. This procedure
is causing loss of information in the Hamiltonian analysis (possibly through a
subtlety in the formulation of the exchange between this $K$-matrix limit and the asymptotic expansion
limit which yields the boundary contributions), giving rise to the observed inconsistencies.
We have thus a strong motivation to systematically search for more general solutions in the SP $A_2^{(1)}$ case. We shall further pursue this significant issue in a separate publication.


\begin{thebibliography}{99}


\bibitem{skl} E.G. Sklyanin, Preprint LOMI E-3-97, Leningrad,
1979.

\bibitem{sts} M.A. Semenov-Tian-Shansky, Funct. Anal. Appl. {\bf
17} (1983) 259.

\bibitem{bv} O. Babelon, C.M. Viallet, Phys. Lett. {\bf B 237} (1990), 411.

\bibitem{avandoikou} J. Avan and A. Doikou, Nucl. Phys. {\bf B800} (2008) 591.

\bibitem{cherednik} I. Cherednik, Theor. Math. Phys. {\bf 61} (1984) 977.

\bibitem{sklyanin} E.K. Sklyanin, Funct. Anal. Appl. {\bf 21} (1987) 164;\\
E.K. Sklyanin, J. Phys. {\bf A21} (1988) 2375.

\bibitem{kulishsklyanin} E.K. Sklyanin; Zap. Nauch. Seminarov LOMI {\bf 95} (1980), 55;\\
P.P. Kulish and E.K. Sklyanin, in Tvarminne Lectures, edited by J.
Hietarinta and C. Montonen, Springer Lectures in Physics {\bf 151}
(1982).

\bibitem{menean} L. Mezincescu and R.I. Nepomechie, Nucl. Phys. {\bf B372} (1992) 597;\\
S. Artz, L. Mezincescu and R.I. Nepomechie, J. Phys. {\bf A28}
(1995) 5131.

\bibitem{durham} P. Bowcock, E. Corrigan, P.E. Dorey and R.H. Rietdijk, Nucl. Phys. {\bf B445} (1995) 469;\\ P. Bowcock, E. Corrigan and R.H. Rietdijk, Nucl. Phys. {\bf B465} (1996) 350.

\bibitem{delius} G. Delius, Phys. Lett. {\bf B444} (1998) 217.

\bibitem{done} A. Doikou and R.I. Nepomechie,  Nucl. Phys. {\bf B521} (1998) 547.

\bibitem{done2}
A. Doikou and R.I. Nepomechie,  Nucl. Phys. {\bf B530} (1998) 641.

\bibitem{gand} G.M. Gandenberger, Nucl. Phys. {\bf B542} (1999) 659;\\
G.M. Gandenberger, \texttt{hep-th/9911178}.

\bibitem{doikousnp} A. Doikou, J. Phys. {\bf A33} (2000) 8797.

\bibitem{dema} G.W. Delius and N. Mackay, Commun. Math. Phys. {\bf 233} (2003) 173.

\bibitem{ann1} D. Arnaudon, J. Avan, N. Cramp\'e, A. Doikou, L. Frappat and E. Ragoucy, J. Stat. Mech. {\bf 0408} (2004) P005.

\bibitem{ann0}
D. Arnaudon, N. Cramp\'e, A. Doikou, L. Frappat and E. Ragoucy, J.
Stat. Mech. {\bf 0502} (2005) P007.

\bibitem{ann2} D. Arnaudon, N. Cramp\'e, A. Doikou, L. Frappat and E. Ragoucy, Int. J. Mod. Phys. {\bf A21} (2006) 1537.

\bibitem{doikounp} A. Doikou, Nucl. Phys. {\bf B725} (2005) 493.

\bibitem{doikouy} A. Doikou, J. Math. Phys. {\bf 46} 053504 (2005).

\bibitem{gama} W. Galleas and M.J. Martins,
Phys. Lett. {\bf A335} (2005) 167.

\bibitem{masa}
R. Malara and  A. Lima-Santos, J. Stat. Mech. {\bf 0609} (2006) P013.

\bibitem{yazh}
W.-L. Yang and Y.-Z. Zhang, JHEP {\bf 0412} (2004) 019.

\bibitem{crdo} N. Cramp\'e and A. Doikou, J. Math. Phys. {\bf 48} 023511 (2007).

\bibitem{paper} A. Doikou, D. Fioravanti and F. Ravanini, Nucl. Phys. {\bf B790} (2008) 465.

\bibitem{doikouatft} A. Doikou, JHEP05 (2008) 091.

\bibitem{ft} L.D. Faddeev, \textsl{Integrable models in $1 + 1$ dimensional
quantum field theory} in `Recent Advances in Field Theory and
Statistical Mechanics' (J.B. Zuber
and R. Stora, Ed.), Les Houches Lectures 1982, North Holland (1984) 561;\\
L.D. Faddeev and L.A. Takhtakajan, {\it Hamiltonian Methods in the
Theory of Solitons}, (1987) Springer-Verlag.

\bibitem{lax} P.D. Lax, Comm. Pure Appl. Math. {\bf 21} (1968)
467.

\bibitem{ZSh} V.E. Zakharov and  A.B. Shabat, Anal. Appl. {\bf 13}
(1979) 13.


\bibitem{AKNS} M.J. Ablowitz, D.J. Kaup, A.C. Newell and H. Segur,
Stud. Appl. Math. {\bf 53} (1974) 249.

\bibitem{abla} M.J. Ablowitz and J.F. Ladik, J. Math. Phys. {\bf
17} (1976) 1011.

\bibitem{maillet} J.-M. Maillet, Phys. Lett. {\bf B162} (1985) 137.

\bibitem{mailfrei2} L. Freidel and J.M. Maillet, Phys. Lett. {\bf B263} (1991) 403.

\bibitem{molev}
G.I. Olshanski, \textsl{Twisted Yangians and infinite-dimensional
classical Lie algebras} in `Quantum Groups' (P.P. Kulish, Ed.),
Lecture notes in Math. {\bf 1510}, Springer (1992) 103;\\  A.I.
Molev, M. Nazarov and G.I. Olshanski, Russ. Math. Surveys {\bf 51}
(1996) 206.

\bibitem{moras} A.I. Molev, E. Ragoucy and P. Sorba, Rev. Math. Phys. {\bf 15} (2003) 789;\\
A.I. Molev, {\it Handbook of Algebra}, Vol. 3, (M. Hazewinkel,
Ed.), Elsevier, (2003), pp. 907.

\bibitem{macnt} A. MacIntyre, J. Phys. {\bf A28} (1995) 1089.

\bibitem{jimbo} M. Jimbo, Commun. Math. Phys. {\bf 102} (1986) 53.

\bibitem{olive} D.I. Olive and N. Turok, Nucl. Phys, {\bf B215}
(1983) 470;\\ D.I. Olive and N. Turok, Nucl. Phys. {\bf B257}
(1985) 277;\\ D.I. Olive and N. Turok, Nucl. Phys. {\bf B265}
(1986) 469.

\bibitem{georgi} H. Georgi, {\it Lie Algebras in Particle Physics}
(Benjamin/Cummings, 1982).

\bibitem{GZ} S. Ghoshal and  A.B. Zamolodchikov, Int. J. Mod.
Phys. {\bf A9} (1994) 3841.

\bibitem{dvg} H.J. de Vega and  A. Gonzalez--Ruiz, Nucl. Phys. {\bf B417} (1994) 553;\\
H.J. de Vega and A. Gonzalez--Ruiz, Phys. Lett. {\bf B332} (1994)
123.

\bibitem{abadrios} J. Abad and M. Rios, Phys. Lett. {\bf B352} (1995) 92.

\bibitem{doma} D. Levy and P.P. Martin, J. Phys. {\bf A27} (1994) L521;\\
A. Doikou and P.P. Martin, J. Phys. {\bf A36} (2003) 2203.

\bibitem{kumu} P. Kulish and A. Mudrov, math/0508289.

\bibitem{nich} A. Nichols, J. Stat. Mech. 0509 (2005) P009.


\end{thebibliography}
\end{document}